\documentclass{article}
\usepackage[margin=1.4in]{geometry}
\usepackage[utf8]{inputenc}

\usepackage{amsmath}
\usepackage{amssymb}
\usepackage{amsthm}
\usepackage{amstext}

\usepackage{array}
\usepackage{graphicx}
\usepackage{wrapfig}
\usepackage{sidecap}
\usepackage{caption}
\usepackage[labelformat=simple]{subcaption}
\usepackage{footnote}

\theoremstyle{plain}

\theoremstyle{definition}

\theoremstyle{remark}

\usepackage{authblk}

\usepackage{biblatex}
\title{The Economics of Automated Market Makers}
\author{Robin Fritsch}
\author{Samuel Käser}
\author{Roger Wattenhofer}
\affil{ETH Zürich\\ \texttt{\{rfritsch,skaeser,wattenhofer\}@ethz.ch}}
\date{}

\begin{document}

\maketitle

\begin{abstract}
This paper studies the question whether automated market maker protocols such as Uniswap can sustainably retain a portion of their trading fees for the protocol. 
We approach the problem by modelling how to optimally choose a pool's take rate, i.e\ the fraction of fee revenue that remains with the protocol, in order to maximize the protocol's revenue.
The model suggest that if AMMs have a portion of loyal trade volume, they can sustainably set a non-zero take rate, even without losing liquidity to competitors with a zero take rate.
Furthermore, we determine the optimal take rate depending on a number of model parameters including how much loyal trade volume pools have and how high the competitors' take rates are.
\end{abstract}

\section{Introduction}

Automated market makers (AMMs) are a paradigm shift in the design of asset exchanges and have gained great popularity for trading cryptocurrencies.
Their combined total trading volume has surpassed \$1 trillion and AMMs are now comparable in size to some of the largest centralized, order book based exchanges \cite{theblock}.

However, in the case of Uniswap, the by far largest AMM protocol by trading volume, currently no trading fees remain with the protocol itself.
In general, it remains an open question whether it is possible for such protocols to retain a part of their fee revenue in a sustainable manner or if they are destined to become common-good infrastructure.
In this paper, we set out to answer this question.

In AMMs, liquidity provider deposit asset reserves into so-called \emph{liquidity pools} to allow other market participants to trade against these reserves.
In return, trading fees paid by traders are distributed pro rata among the liquidity providers.
However, AMM protocols can decide to keep a certain portion of these fees. We call the fraction of fee revenue that is kept by the protocol the \emph{take rate}. Uniswap currently has a take rate of zero since all its fees go to liquidity provider. On the other hand, Sushiswap's take rate (one of Uniswap's largest competitors) is 16.7\% \cite{medium_feeswitch}.
While a higher take rate promises the protocol a higher revenue, it also reduces the return on investment (ROI) for liquidity providers and will incentivize them to move to competing pools, thereby taking trade volume with them.
This aspect of competition is particularly relevant in the DeFi space where new competitors can emerge from a protocol's code being forked.
In this paper, we define a model for the problem, and quantify this tradeoff in order to find the optimal take rate.

As a quick thought experiment, consider two pools that are perfectly balanced (i.e.\ have an equal ratio of reserves) and charge the same trading fees.
If all trades are executed optimally through the two pools, trading volume will be split proportional to the pool sizes between the two pools (see the calculations in Appendix \ref{app:optimal_trade} based on \cite{fritsch21optimalfees}).
Hence, the ROI of liquidity providers in the two pools will be equal before considering the pools' take rates, regardless of the exact liquidity distribution. 
After taking the take rates into account, liquidity providers in the pool with the lower take rate will always have a higher ROI. This would incentivize all liquidity to move to the pool with the lower take rate.
So under these (quite naive) assumptions, the pool with the lowest take rate will always attract all liquidity and subsequently all trade volume.
In particular, any pool that introduces a non-zero take rate must fear the creation of a forked pool with a zero take that would attract all liquidity.

These considerations would suggest that AMM protocols are destined to become common-good infrastructure and are not able to extract any value from the trading fees for themselves.
But will this turn out to be true in practice? How realistic are these simple assumptions? After all, AMMs with non-zero take rates do exist in practice.
In this paper, we design a more sophisticated model to also take into account that market participants do not always act completely optimally:
For traders as well as liquidity providers, we allow for a fraction of them to favor a certain pool.
With this model, we are able to evaluate how pools should optimally set their take rates and if they are able to sustainably extract revenue for the protocol.

First, we use the model to arrive at a number of analytical results. In a second step, we add more details such as arbitrage trading to the model and run simulations with it.

We find that there is a clear optimal take rate for the pool that maximizes the protocol revenue in many cases.
Our model suggest, that if a fraction of the traders in a pool is loyal to the pool (and does not trade optimally), the pool can sustainably set a non-zero take rate, even without losing liquidity to competitors with a zero take rate.
Furthermore, our results determine the optimal take rate depending on a number of model parameters including how much loyal trade volume pools have and how high the competitors' take rates are.

\subsection{Background}

Automated market maker protocols are smart contracts deployed on a blockchain that allow traders to exchange one asset for another in a fully noncustodial manner.
On a high level, they work as follows:
First a liquidity pool for a certain set of assets is created, and liquidity providers (LPs) deposit reserves of these assets into the pool.
Then traders can then swap these assets against the pool. The exchange rate of a trade is determined by the pool-specific trade function depending on the reserves currently in the pool. Uniswap and most other AMMs are so-called constant product market makers, meaning the trade function ensures that the product of the reserves in the pool stays constant.

For every trade, a percentage of the trade size is paid as a trading fee. These fees are (for the most part) distributed pro rata among liquidity providers in the pool.
However, protocols can decide to keep a portion of the fees. 
More precisely, a part of the fees may be diverted to the protocol's treasury or be distributed among the holders of the protocol's governance token which include the team, investors and other stakeholders.
While other AMMs such as Sushiswap and Pancakeswap already do this, Uniswap currently does not \cite{medium_feeswitch}. This however, can be changed by a governance decision, i.e.\ a vote by the governance token holders. For the version 2 Uniswap contract (v2), governance has the option to introduce a take rate to 16.7\% \cite{uniswapv2} while for Uniswap v3, a take rate between 10\% and 25\% can be set\cite{uniswapv3}.
This decision which is often referred to as the \emph{fee switch} has been eagerly discussed among the Uniswap community\footnote{\url{https://gov.uniswap.org/t/temperature-check-fee-switch-v2-should-be-turned-on/13537/19},\\\url{https://gov.uniswap.org/t/temperature-check-ultrasound-uni-fee-switch-organization-funding/16119/38}}.

\subsection{Related Work}

The study of mechanisms to automate the process of providing liquidity to a market dates back at least two decades.
An early such automated market maker is Hanson's logarithmic market scoring rule (LMSR) which was proposed for use in prediction markets \cite{hanson03}.

Most AMMs that have recently become popular in Decentralized Finance (DeFi) for trading cryptocurrencies however, are of a new type called \emph{constant function market maker} (CFMM) \cite{angeris20oracles}.
More detailed studies of AMMs including how they behave alongside traditional, order book based exchanges can for instance be found in \cite{angeris21uniswap} and \cite{aoyagi2021liquidity}.
For an overview of the emerging area of Decentralized Finance of which AMMs are only one part, we refer to \cite{werner21sok}.

In order to simulate the trade flow, we use considerations from \cite{fritsch21optimalfees} on optimal trading. The more general problem of optimally routing trades through a network of AMMs has been studied by Danos et al.\ \cite{danos21routing} and Angeris et al.\ \cite{angeris22routing}.

An economic study of a different aspect of the decentralized exchanges was recently performed by Zhang et al.\ \cite{zhang22economic}.
The paper analyzes how competing decentralized exchanges can optimally reward traders with governance tokens (so-called transaction fee mining).

\section{Model}

We study the problem of setting the optimal take rate for AMM pools that compete with each other for liquidity and volume. 
More precisely, we model two competing AMM pools, and optimize the take rate for one of the pools.
For an adjustment of the take rate, we model how liquidity and trade volume reacts, and how this affects the trading fees earned by the pool.
The goal is to find the optimal take rate that maximizes the protocol revenue generated from the pool.

Let the sizes of the two pools be denoted by $L_1$ and $L_2$, respectively.
We optimize the take rate of pool 1 while it competes with pool 2, the \emph{competitor pool}.
We assume that the two pools charge the same trading fee $f$.
Furthermore, let $V$ be the total trade volume that is traded in the two pools.
Lastly, we denote the take rates of the two pools by $t_1$ and $t_2$ where $0\leq t_1, t_2\leq 1$. They indicate that a fraction of $t_i$ of the trading fees in pool $i$ goes to the protocol, while a fraction of $1-t_i$ is distributed among liquidity providers.

Generally speaking, we make the assumption that liquidity providers and traders act rationally, and that the market is efficient.
More specifically, for traders, we assume that they optimally route their trades through the pools to maximize the output of their trade.
For liquidity providers, we assume that they will move their liquidity if they can achieve a higher return on investment from trading fees in another pool.
This means an equilibrium in the liquidity distribution is reached when the return on investment is equal in both pools. In such a situation, no liquidity provider has an incentive to move pools.

As mentioned in the introduction, these naive assumptions would imply that the pool with the lowest take rate can attract all liquidity and trade volume.
This very simple model however, does not lead to realistic results as the existence of AMMs with non-zero take rates shows. In reality, not all market participants act perfectly optimal.
In particular for traders, it has been observed empirically that many trades are not executed optimally even among established pools \cite{fritsch21concentratedliquidity}. For an empirical study of trade routing, see \cite{berg22empirical}.

Furthermore, there are a number of practical obstacles to a perfectly efficient market.
Not all traders and liquidity providers have knowledge of all pools, especially not recently created ones.
Moreover, traders and liquidity providers that are aware of the new pools might view them as more risky than long-established pools.
To take these effects into account and make the model more realistic, we extend our model in the two following ways.

\subsection{Sticky Trade Volume}

We model the fact that not all trades are routed optimally by introducing a so-called \emph{sticky rate $s_i$} for each pool.
The sticky rate indicates that a fraction of $s_i$ of trade volume will be executed entirely in pool $i$ and not be routed optimally through both pools.

More precisely, the two pools have the sticky rates $s_1$ and $s_2$, respectively, where $1\leq s_1,s_2\leq 1$ and $s_1+s_2\leq 1$.
In the simulation, we also take into account possible arbitrage trades that result from such suboptimal trading.

Of course, there are limits to the stickiness of trades: If a pool's price deviate too strongly from the market, even sticky traders will likely consider alternatives.
To account for this, we add an extra condition for the simulations: If the price a sticky trader would get in their desired pool is a lot worse than the optimal price, the trade will be routed optimally instead.
As a threshold for the price difference, we choose 10\%.

\subsection{Sticky Liquidity}

The assumption that liquidity providers act completely optimal is also not entirely realistic.
They likely prefer a well-known and trusted pool over a new and unknown competitor, and may accept a lower return on their liquidity in the trusted pool.
To take this into account, we also add the notion of \emph{sticky liquidity} to our model.
This introduces a difference $d$ in ROI which LPs accept on their liquidity in the original pool.
More precisely, liquidity providers will accept if the potential return on investment in the competitor pool is $(1+d)$-times as high as in the current pools without moving their liquidity.
In other words, an equilibrium in the liquidity distribution is reached if the return on investment in pool 2 is $(1+d)$-times as high as in pool 1.
Note that by setting $d=0$ the case that the liquidity provision market is completely efficient and LPs do not prefer one pool over another is modeled.

\subsection{Assumptions}

In the following, we briefly summarize and justify the assumptions we make in our model.

First, we model two competing pools. This already captures the vital aspect of competition between pools.
The case of three or more pools is left to future work.

We also assume that the two competing pools use the same trading fee. It is possible to choose differing fees in our model, but in order to isolate the effects of different take rates, we choose the trading fee to be equal in both pools.

Furthermore, we make the following assumptions on trading behavior.
Note that we do not assume all trades are routed optimally, since we introduce sticky traders for both pools.
For the analytical part, we assume that the optimally routed trade volume is split among the pool proportionally to pool sizes.
This is true for two balanced CPMM pools \cite{fritsch21optimalfees}.
Of course, the prices in pools can occasionally slightly differ.
But since the imbalance can go in both directions, it is reasonable to assume that these effects average out in the long run.
More importantly, we do calculate the exact optimal splits in our simulations and take imbalances in the pools into account. We also consider possible arbitrage trading when a price difference between the pool occurs. This way, we can assess the effect of our assumption in the analytical part.

In our simulations, we assume both pools use a constant product market maker (CPMM) mechanism for calculating the optimal trade routing. This is currently the most used mechanism for AMM. The largest DEX, version 3 of Uniswap, is a exception to a small extent as adds the concept of concentrated liquidity to a CPMM \cite{uniswapv3}.
However, it still behaves like a CPMM for small price changes.
Hence, we see this assumption as very reasonable.

Finally, we highlight two slightly implicit assumptions: We assume that the total trading volume $V$ and the total liquidity in the market, i.e. $L_1+L_2$, are constant and do not depend on the take rates set by the pools.

\section{Analytical Approach}

We begin by finding the optimal take rate analytically our model.
We later run simulations with a number of additions to the model and compare the results in order to analyze potential differences.

Remember that a fraction $s_i$ of the total volume $V$ is sticky volume that will be traded in pool $i$.
The remaining volume $(1-s_1-s_2)V$ is traded optimally. Based on the calculations in Appendix \ref{app:optimal_trade}, we assume this volume is split proportionally to the pool sizes. Hence, the volume executed in pool $i$ is
\begin{equation}\label{eq:pool_volume}
    v_i = s_i V + (1-s_1-s_2)\frac{L_i}{L_1+L_2}V
\end{equation}
for $i=1,2$.
On the trade volume, traders pay a trading fee $f$, and of this fee revenue, a fraction of $t_I$ goes to the protocol while $1-t_i$ stays with the liquidity providers.
Hence, the ROI of liquidity providers in pool i equals
\begin{equation}\label{eq:pool_roi}
    r_i = \frac{(1-t_i)v_i f}{L_i}.
\end{equation}
According to the sticky liquidity assumption, liquidity providers accept a difference $d$ between the ROIs of the two pools.
More precisely, we assume an equilibrium in the liquidity distribution between the two pools is reached if
\begin{equation}\label{eq:liquidity_equilibrium}
    r_1 (1+d) = r_2.
\end{equation}
In particular, when choosing $d=0$, we assume that LPs have no preference for one of the pools, and that in the equilibrium the ROI of LPs is equal in both pool.

By plugging \eqref{eq:pool_roi} and \eqref{eq:pool_volume} into \eqref{app:liquidity_equilibrium} it can be solved for the distribution of liquidity in the equilibrium.
More precisely, we can determine the fraction $l_1:=L_1/(L_1+L_2)$ of liquidity in pool 1 in the equilibrium.
The detailed derivations can be found in Appendix \ref{app:take_rate}.
They yield
\begin{equation*}
    l_1 = \begin{cases}
        \frac{p}{2}+\sqrt{\frac{p^2}{4}-q} & \text{if } 1-t_1 > \frac{1-t_2}{1+d} \\
        \frac{p}{2}-\sqrt{\frac{p^2}{4}-q} & \text{if } 1-t_1 < \frac{1-t_2}{1+d}.
    \end{cases}
\end{equation*}
where
\begin{align*}
    p &= 1+\frac{(1+d)(1-t_1)s_1+(1-t_2)s_2}{(1-s_1-s_2)((1-t_2)-(1+d)(1-t_1))}\\
    q &= \frac{(1+d)(1-t_1)s_1}{(1-s_1-s_2)((1-t_2)-(1+d)(1-t_1))}.
\end{align*}
With the fraction of liquidity in pool 1, we are able to calculate the volume in pool 1 according to \eqref{eq:pool_volume}, and with this the revenue of the protocol.
Remember that the protocol receives a fraction of $t_1$ of trading fees. So the protocol revenue of pool 1 equals
\begin{equation*}
    t_1v_1 f = t_1 (s_1 + (1-s_1-s_2)l_1) Vf.
\end{equation*}
The goal is to find the $t_1$ that maximizes this expression, more precisely, the term
\begin{equation*}
    \mathrm{rev}_1 := t_1 (s_1 + (1-s_1-s_2)l_1)
\end{equation*}
since $V$ and $f$ are constant in our model.

\subsection{Results}

In the following, we discuss our results for a number of model configurations in more detail.
We consider two main scenarios:
\begin{enumerate}
    \item[(1)] The competitor pool is a newly-created fork with a take rate of zero and sticky rate of zero.
    \item[(2)] The competitor pool is an established competitor with a non-zero take rate and possibly a non-zero sticky rate.
\end{enumerate}

In the first scenario, and more generally when $s_2=0$ holds, the expressions for $l_1$ and the protocol revenue $rev_1$ can be significantly simplified.
The detailed calculations can be found in Appendix \ref{app:take_rate}. They show that, in this case,
\begin{equation*}
    l_1 = \begin{cases}
        1 & \text{if } 1-t_1 \geq \frac{1}{1+d}(1-s_1)(1-t_2) \\
        p-1 & \text{if } 1-t_1 \leq \frac{1}{1+d}(1-s_1)(1-t_2).
    \end{cases}
\end{equation*}
and
\begin{equation*}
    \mathrm{rev}_1 = \begin{cases}
        t_1 & \text{if } 1-t_1 \geq \frac{1}{1+d}(1-s_1)(1-t_2) \\
        \frac{s_1(1-t_2)}{(1+d)-(t_2+d)/t_1} & \text{if } 1-t_1 \leq \frac{1}{1+d}(1-s_1)(1-t_2).
    \end{cases}
\end{equation*}

\begin{figure}[ht]
\begin{subfigure}{0.50\textwidth}
\centering
\includegraphics[width=\linewidth]{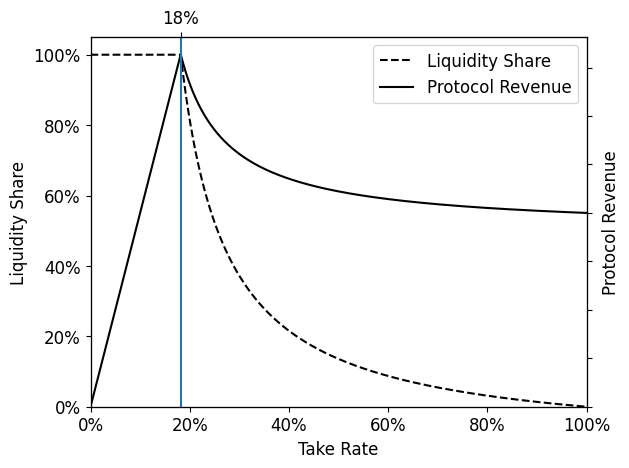}
\caption{$t_2 = 0.0, s_1 = 0.1, s_2 = 0.0, d = 0.1$}
\label{fig:forked_d_0.1}
\end{subfigure}%
\begin{subfigure}{0.50\textwidth}
\centering
\includegraphics[width=\linewidth]{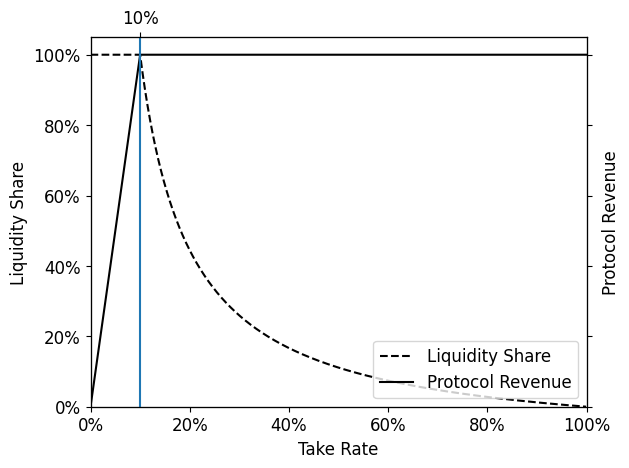}
\caption{$t_2 = 0.0, s_1 = 0.1, s_2 = 0.0, d = 0.0$}
\label{fig:forked_d_0.0}
\end{subfigure}
\caption{The share of liquidity and the protocol revenue of pool 1 for different take rates of pool 1 when competing against a forked competitor (zero take rate and zero sticky rate).}
\label{fig:forked_competitor}
\end{figure}

Figure \ref{fig:forked_d_0.1} plots these functions depending on $t_1$ for $t_2=0.0, s_1=0.0, s_2=0.0$ and $d=0.1$.
Note that for $\mathrm{rev}_1$ the upper term is strictly increasing in $t_1$ while the lower one is strictly decreasing. Hence, the function attains its maximum value for $1-t_1=(1-s_1)(1-t_2)/(1+d)$.
This is also the point up to which pool 1 can increase its take rate without losing any liquidity to a forked competitor pool.
In other words, the optimal take rate for pool 1 is
\begin{equation}\label{eq:opt_take_rate}
    t_1^*=1-\frac{1}{1+d}(1-s_1)(1-t_2).
\end{equation}
Furthermore, since the forked pool does not charge a take rate, i.e.\ $t_2=0$, this simplifies to $t_1^*=1-(1-s_1)/(1+d)$. 
For $d=0$, i.e.\ we look for the equilibrium where the return of investment is equal in both pools (``no sticky liquidity''), this further simplifies to $t_1=s_1$.
So in this case, a protocol should set its take rate as high as it estimates its sticky rate to be.

Two situations of a pool competing with forks are shown in Figure \ref{fig:forked_competitor}. In both cases, pool 1's sticky rate is 10\%, and Figure \ref{fig:forked_d_0.1} additionally features an accepted ROI difference of 10\%.
The protocol revenue displays a clear peak in Figure \ref{fig:forked_d_0.1} indicating the clear optimal take rate.
For Figure \ref{fig:forked_d_0.0}, a take rate of 10\% is the best choice since the liquidity in the pool decreases rapidly for larger take rates even though the protocol revenue does not.
It is also important to note that the right part of the plots where the take rate in unrealistically high and the fraction of liquidity in the pool is very low, must be taken with a grain of salt. It is less relevant for finding the optimal take rate.

In general, note that in this scenario, the optimal take rate $t_1^*$ increases linearly in the pool's sticky rate $s_1$ and its competitor's take rate $t_1$.
In particular, setting $d=0$ in \eqref{eq:opt_take_rate} yields
\begin{equation*}
t_1^*=t_2+s_1(1-t_2).
\end{equation*}

\begin{figure}[ht]
\centering
\includegraphics[width=0.50\textwidth]{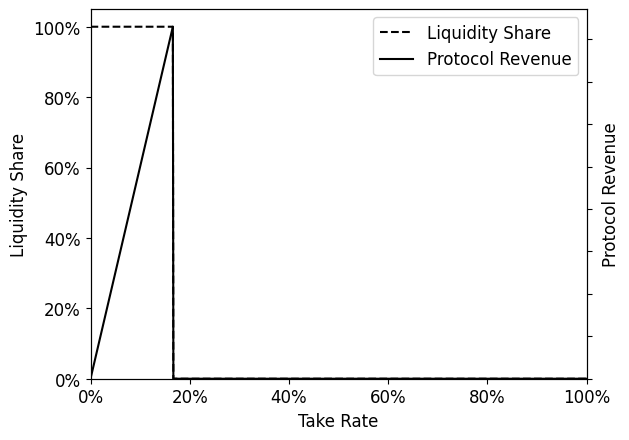}
\caption{Scenario without sticky volume or sticky liquidity, more precisely $t_2 = 0.167, s_1 = 0.0, s_2 = 0.0, d = 0.0$.}
\label{fig:no_sticky}
\end{figure}

As a sidenote, consider the situation where there is no sticky trade volume and no sticky liquidity, i.e.\ $s_1=0.0,s_2=0.0$ and $d=0.0$.
An example of this is shown in Figure \ref{fig:no_sticky}.
Then the model yields the expected result (as discussed in the introduction): The pool with the lower take rate attracts all liquidity and volume.
\newline

\begin{figure}[ht]
\begin{subfigure}{0.50\textwidth}
\centering
\includegraphics[width=\linewidth]{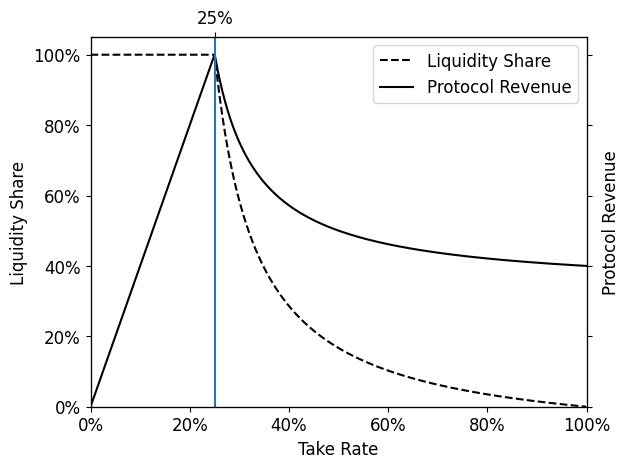}
\caption{$t_2 = 0.167, s_1 = 0.1, s_2 = 0.0, d = 0.0$}
\label{fig:established_s2_0.0}
\end{subfigure}%
\begin{subfigure}{0.50\textwidth}
\centering
\includegraphics[width=\linewidth]{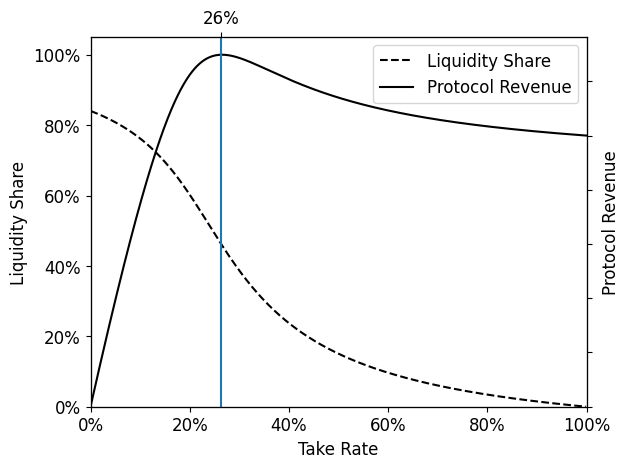}
\caption{$t_2 = 0.167, s_1 = 0.1, s_2 = 0.05, d = 0.0$}
\label{fig:established_s2_0.05}
\end{subfigure}
\caption{The share of liquidity and the protocol revenue of pool 1 for different take rates of pool 1 when competing against an established competitor with 16.7\% take rate (like Sushiswap).}
\label{fig:established_competitor}
\end{figure}

In the second scenario, more precisely when $s_2\neq 0$, things are slightly more complicated.
Then it is no longer as simple to find a closed expression for $t_1^*$.
However, it is nevertheless possible to calculate the $t_1$ that maximizes the protocol revenue numerically.
Two examples of this scenario are shown in Figure \ref{fig:established_competitor}. In both cases, the pool's sticky rate is 10\% and the competing pool charges a take rate of 16.7\% (like Sushiswap).
For Figure \ref{fig:established_s2_0.05}, we additionally assume a sticky rate of 5\% for the competitor pool.
This leads to the notable (and arguably more realistic) difference that the pool can never attract all liquidity from the competitor pool by setting a low take rate.
Note that in this case the choice of a take rate is not 100\% obvious.
In the previous three cases, the choice for the take rate was relatively clear: The protocol revenue maximizing take rate coincided with the maximum take rate for which the pool still attracts all liquidity.
From Figure \ref{fig:established_s2_0.05} the choice is more of a judgement call.
With a take rate of 26\% the protocol revenue is maximized. However, then the pool attracts less than 50\% of liquidity. For a 20\% take rate on the other hand, the protocol still receives 94\% of the maximally possible revenue while the pool attracts about 60\% of liquidity.


\section{Simulations}\label{sec:simulations}

In the following, we will run a number of simulations in order to make our model and results emulate the real world more closely.
We run a simulation for all scenarios considered in the analytical part which allows us to assess the effects of some of the assumptions made in that part.

Our model includes a number of parameters, namely the sticky rates $s_1$ and $s_2$ of the pools, the competitor pool's take rate $t_2$ and the accepted ROI difference $d$.
For a given set of these parameters, the goal is to find the optimal take rate $t_1$ of pool 1.
To achieve this, we run a simulation for every take rate $t_1$ of pool 1 in discrete steps of size 1\% and calculate the protocol revenue each time.

To run such simulations, it is necessary to choose values for the available liquidity and the trades that are to be executed.
For this, we use real world data in the form of Uniswap trade data. We choose to consider the ETH-USDC pair since it is the most traded pair on Uniswap\footnote{\url{https://info.uniswap.org/#/}, \url{https://v2.info.uniswap.org/pairs}}.
As total amount of liquidity, we use the amount of liquidity in the Uniswap v2 ETH-USDC pool at the beginning of August 2021.
This is the liquidity we assume to be shared between the two competing pools we simulate.
Furthermore, we use the next 10.000 trades in the Uniswap v2 ETH-USDC pool after this point in time.

The base part of the simulation is the \emph{trade simulation}.
This simulates all trades in the pools for fixed pool sizes:
In the beginning the pools are set to be balanced, i.e.\ to have the same marginal price. Remember that both pools are assumed to use the CPMM mechanism.
Then the execution of all trades is simulated.
As sticky volume, we choose the smallest trades in the sample. (These are most likely to be sticky as they can profit least from routing optimally.)
More precisely, we choose the smallest trades that make up a fraction of $s1+s2$ of trade volume to be sticky.
Among these we randomly select a fraction of $s1$ to be sticky in pool 1, while the remaining trades are sticky in pool 2.
All trades that are not selected to be sticky are routed optimally.
In case a trade is chosen to be a sticky trade in one of the pools, we check if the price the trader will get in that pool differs from the optimal price by a certain threshold (10\% in our simulations). If so, the trade is routed optimally instead.
After each trade, we check whether an arbitrage trade is possible between the two pools and, if it is, execute the optimal arbitrage trade.

For fixed pools sizes, this trade simulation allows us to find the fee revenues and ROIs of the liquidity providers in both pool.
To now find the liquidity equilibrium, i.e.\ the distribution of liquidity between the pools which is described by equation \eqref{eq:liquidity_equilibrium} in the analytical part, we discretize the space of liquidity distributions. We do so by choosing discrete steps of size 0.5\% for the fraction of liquidity in pool 1.
We then run the trade simulation for every liquidity distribution and find the one where $r_1(1+d)=r_2$ holds.

For the equilibrium liquidity distribution, the fee revenue of pool 1 and subsequently the protocol revenue can be determined by another run of the trade simulation.
By running the whole simulation for different $t_1$, we can plot the protocol revenue against the take rate and find the optimal take rate $t_1^*$.
Examples of this can be seen in Figure \ref{fig:simulations}.


\subsection{Results}

\begin{figure}
\begin{subfigure}{0.50\textwidth}
\centering
\includegraphics[width=\linewidth]{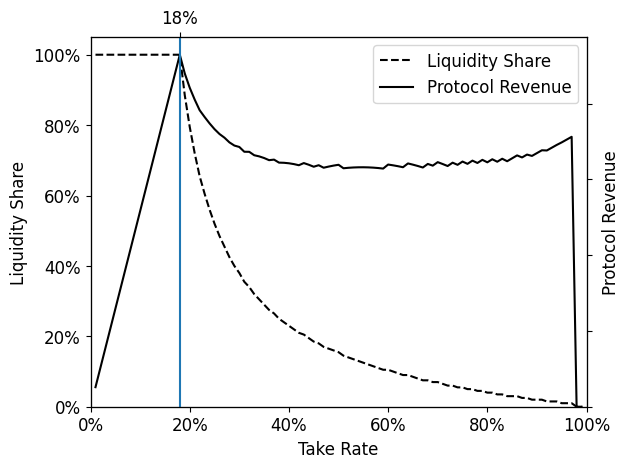}
\caption{$t_2 = 0.0, s_1 = 0.1, s_2 = 0.0, d = 0.1$}
\label{fig:sim_forked_d_0.1}
\end{subfigure}%
\begin{subfigure}{0.50\textwidth}
\centering
\includegraphics[width=\linewidth]{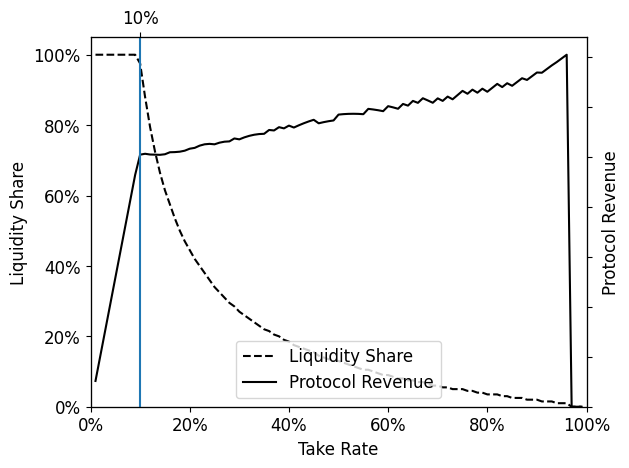}
\caption{$t_2 = 0.0, s_1 = 0.1, s_2 = 0.0, d = 0.0$}
\label{fig:sim_forked_d_0.0}
\end{subfigure}

\begin{subfigure}{0.50\textwidth}
\centering
\includegraphics[width=\linewidth]{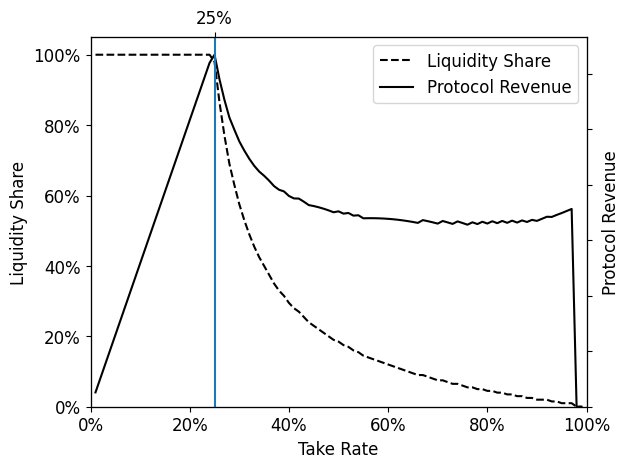}
\caption{$t_2 = 0.167, s_1 = 0.1, s_2 = 0.0, d = 0.0$}
\label{fig:sim_established_s2_0.05}
\end{subfigure}%
\begin{subfigure}{0.50\textwidth}
\centering
\includegraphics[width=\linewidth]{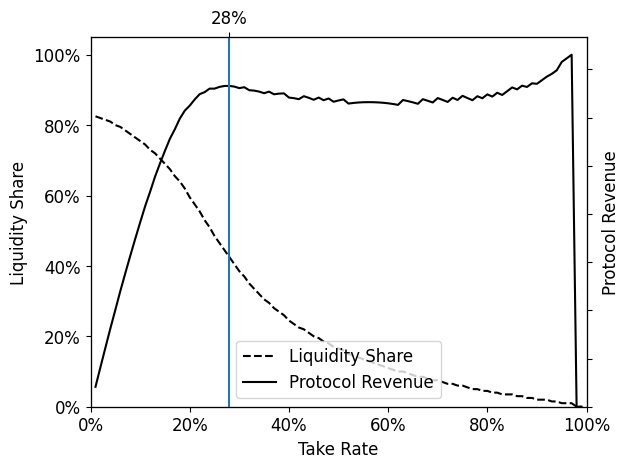}
\caption{$t_2 = 0.167, s_1 = 0.1, s_2 = 0.05, d = 0.0$}
\label{fig:sim_established_s2_0.0}
\end{subfigure}

\caption{Simulation results for four scenarios: The upper two are competing with a forked pool (zero take rate), the lower two are competing with an established pool (16.7\% take rate).}
\label{fig:simulations}
\end{figure}

When there is no sticky volume and liquidity, the simulation leads to the exact same plot and results as the analytical part (see Figure \ref{fig:no_sticky}).
The simulation results for all four remaining scenarios discussed in the previous section are shown in Figure \ref{fig:simulations}.
The two upper plots show the situation with a forked competitor with zero take rate and zero sticky rate. In particular, these are the same scenarios as shown in Figure \ref{fig:forked_competitor}.
We find that the simulations lead to very similar results as the analytical part.
Note that the sudden drop in protocol revenue for take rate near to 100\% stems from the fact that the fraction of liquidity in pool 1 is then rounded to zero.
In general, the right half of the plots is less relevant for finding the best take rate since the take rate is unrealistically high there, and the fraction of liquidity in the pool drops to very low and undesirable levels.

One slight difference that can be seen between the simulation and the analytical results, is that the simulations tend to estimate the protocol revenue slightly higher than the analytical results for large take rates (e.g.\ compare Figures \ref{fig:forked_d_0.0} and \ref{fig:sim_forked_d_0.0}).
This however does not greatly affect the choice of the optimal take rate much.
For instance, the most desirable take rate is arguably still 10\% based on Figure \ref{fig:sim_forked_d_0.0} when also taking the liquidity share into account.

On closer investigation, the slight difference in the simulation results can be explained as follows.
Sticky trades create price differences between the two pools which are closed by subsequent optimal trades or arbitrage trades.
If they are closed by following trades, this means that a part of the non-sticky volume is not split optimally. Instead, the split depends on the direction of the trade.
In the case of an arbitrage trades, both pools receive the same amount of trade volume.
If one pool is small compare to the other, this part of volume is distributed more equally among the pools than the optimal trades would be.
Hence, the simulations find sightly higher fee and protocol revenues for the smaller pool.

When simulating an established competitor with a non-zero take rate, we come to the same conclusions:
The lower two plots of Figure \ref{fig:simulations} show the same as scenarios as Figure \ref{fig:established_competitor}. The analytical and simulation plots are of the same shape, and in particular, the choices of optimal take rates from the plots are very similar.

\section{Conclusion}

We see this paper as a first step towards optimizing protocol revenues of automated market makers in particular, and a study of their economics in general.
We suggest a first model for the problem and get an impression of optimal take rates and how they depend on different parameters.
Of course, the model we suggest does not capture all aspects of reality, and can be improved in future work.
An interesting direction could be to conduct empirical studies on how liquidity providers react to changing take rates of pools.

Furthermore, a next step could be to consider Nash equilibria of take rates:
If one pool sets an optimal take rate according to the current state of the market, as described in this paper, the situations for other pools changes.
This will lead to them adjusting their take rates.
An equilibrium is reached when the current take rate is optimal for every pool.

\printbibliography

\appendix

\section{Finding the Liquidity Equilibrium}\label{app:liquidity_equilibrium}

This section describes how the equilibrium in the liquidity distribution can be derived from the equality $r_1 (1+d) = r_2$.
Inserting the term for the ROIs and trade volumes (i.e. equations \eqref{eq:pool_roi} and \eqref{eq:pool_volume}), yields
\begin{align*}
    (1+d)(1-t_1)\left(\frac{s_1}{L_1} + \frac{1-s_1-s_2}{L_1+L_2}\right) Vf &= (1-t_2)\left(\frac{s_2}{L_2} + \frac{1-s_1-s_2}{L_1+L_2}\right)Vf\\
    \Leftrightarrow\,\,
    (1+d)(1-t_1)s_1\frac{1}{L_1} - (1-t_2)s_2\frac{1}{L_2} &= (1-s_1-s_2)\left((1-t_2)-(1+d)(1-t_1)\right)\frac{1}{L_1+L_2}.
\end{align*}
Multiplying by $L_1+L_2$ and substituting $l_1=L_1/(L_1+L_2)$, leads to
\begin{align*}
    (1+d)(1-t_1)s_1\frac{1}{l_1} - (1-t_2)s_2\frac{1}{1-l_1} &= (1-s_1-s_2)\left((1-t_2)-(1+d)(1-t_1)\right)\\
    \Leftrightarrow\,\,
    (1+d)(1-t_1)s_1-((1+d)(1-t_1)s_1 &+ (1-t_2)s_2)l_1 \\
    &= (1-s_1-s_2)\left((1-t_2)-(1+d)(1-t_1)\right)\left(l_1-l_1^2\right).
\end{align*}
In the second step above, the equation was multiplied by $l_1(1-l_1)$.
The latter equation can be rewritten as a quadratic equation in $l_1$ of the form $l_1^2-pl_1+q=0$ where the coefficients are
\begin{align*}
    p &= 1+\frac{(1+d)(1-t_1)s_1+(1-t_2)s_2}{(1-s_1-s_2)((1-t_2)-(1+d)(1-t_1))}\\
    q &= \frac{(1+d)(1-t_1)s_1}{(1-s_1-s_2)((1-t_2)-(1+d)(1-t_1))}.
\end{align*}
The solution of the quadratic equation is then given by
\begin{equation*}
    l_1 = \frac{p}{2} \pm \sqrt{\frac{p^2}{4}-q}.
\end{equation*}
We briefly confirm that the $p^2/4-q$ term is non-negative and we get rational solutions. This can be seen by substituting $T:=(1-s_1-s_2)((1-t_2)-(1+d)(1-t_1))$ and rewriting the expression:
\begin{align*}
    \frac{p^2}{4}-q &= \frac{1}{4T^2}\left((T+(1+d)(1-t_1)s_1+(1-t_2)s_2))^2 - 4T(1+d)(1-t_1)s_1\right)\\
    &= \frac{1}{4T^2}\left((T-(1+d)(1-t_1)s_1+(1-t_2)s_2))^2 + 4(1+d)(1-t_1)s_1(1-t_2)s_2\right)
\end{align*}

So the quadratic equation has one or two solutions. 
Next, we determine which solution is valid in our context, as we expect $0\leq l_1\leq 1$.
To do so, we distinguish the following two cases. (Note that the expression is not defined for $(1-t_2)-(1+d)(1-t_1)=0$.)

\textbf{1st case:} $(1-t_2)-(1+d)(1-t_1)<0$. Then $q<0$, $p<1$ and $p-1<q$.
Therefore
\begin{equation*}
    \frac{p}{2}-\sqrt{\frac{p^2}{4}-q}<\frac{p}{2} - \left|\frac{p}{2}\right|\leq 0
\end{equation*}
is not a valid solution.
On the other hand,
\begin{equation*}
    \frac{p}{2}+\sqrt{\frac{p^2}{4}-q}>\frac{p}{2} + \left|\frac{p}{2}\right|\geq 0.
\end{equation*}
Furthermore,
\begin{align*}
    \frac{p}{2}+\sqrt{\frac{p^2}{4}-q}\leq 1 \quad &\Leftrightarrow\quad
    \sqrt{\frac{p^2}{4}-q}\leq 1-\frac{p}{2}\\
    &\Leftrightarrow\quad \frac{p^2}{4}-q \leq 1-p+\frac{p^2}{4} \quad\text{and}\quad 1-\frac{p}{2}\geq 0\\
    &\Leftrightarrow\quad p-1\leq q \quad\text{and}\quad 1-\frac{p}{2}\geq 0.
\end{align*}

The latter is true. So in this case, $l_1=p/2+\sqrt{p^2/4-q}$ is the only valid solution.

\textbf{2nd case:} $(1-t_2)-(1+d)(1-t_1)>0$. In this case, $q>0$, $p>1$ and $p-1>q$.
Thus,
\begin{equation*}
    \frac{p}{2}-\sqrt{\frac{p^2}{4}-q}>\frac{p}{2} - \left|\frac{p}{2}\right|\geq 0.
\end{equation*}
Since we also have
\begin{align*}
    \frac{p}{2}-\sqrt{\frac{p^2}{4}-q}\leq 1 \quad\Leftrightarrow\quad \frac{p}{2}-1\leq \sqrt{\frac{p^2}{4}-q} \quad\Leftarrow\quad \frac{p^2}{4}-p+1\leq \frac{p^2}{4}-q \quad\Leftrightarrow\quad q\leq p-1,
\end{align*}
the solution $l_1=p/2-\sqrt{p^2/4-q}$ is valid. The second solution however, is not, as
\begin{align*}
    \frac{p}{2}+\sqrt{\frac{p^2}{4}-q}\leq 1 \quad\Leftrightarrow\quad \sqrt{\frac{p^2}{4}-q}\leq 1-\frac{p}{2} \quad\Rightarrow\quad \frac{p^2}{4}-q\leq \frac{p^2}{4}-p+1 \quad\Leftrightarrow\quad p-1\leq q.
\end{align*}

To summarize, we arrive at the following solution
\begin{equation*}
    l_1 = \begin{cases}
        \frac{p}{2}+\sqrt{\frac{p^2}{4}-q} & \text{if } 1-t_1 > \frac{1}{1+d}(1-t_2) \\
        \frac{p}{2}-\sqrt{\frac{p^2}{4}-q} & \text{if } 1-t_1 < \frac{1}{1+d}(1-t_2).
    \end{cases}
\end{equation*}

\section{Trade Revenue Maximum}\label{app:take_rate}

In the following, we discuss the trade revenue function in the case that $s_2=0$.
Then $q=p-1$ and we can simplify
\begin{equation*}
    \sqrt{\frac{p^2}{4}-q} = \sqrt{\left(\frac{p}{2}-1\right)^2} = \left|\frac{p}{2}-1\right|.
\end{equation*}
To further simplify $l_1$, we need to know whether $p/2-1$ is positive or negative.
In the case that $(1-t_2)-(1+d)(1-t_1)<0$, we have $p<1$ and, in particular, $p/2-1<0$. Therefore,
\begin{equation*}
    l_1 = \frac{p}{2} + \left|\frac{p}{2}-1\right| = \frac{p}{2} + \left(1-\frac{p}{2}\right) = 1.
\end{equation*}
On the other hand, if $(1-t_2)-(1+d)(1-t_1)>0$, there are two possible cases depending on whether $p/2-1$ is positive or negative. Notice that
\begin{align*}
    \frac{p}{2}-1\geq 0 \quad\Leftrightarrow\quad \frac{(1+d)(1-t_1)s_1}{(1-s_1)((1-t_2)-(1+d)(1-t_1))} \geq 1
    \quad\Leftrightarrow\quad (1+d)(1-t_1)\geq (1-s_1)(1-t_2).
\end{align*}
If these inequalities hold, we again find that $l_1 = 1$. Otherwise, we have $l_1=p-1$.
Putting everything together, gives us
\begin{equation*}
    l_1 = \begin{cases}
        1 & \text{if } 1-t_1 \geq \frac{1}{1+d}(1-s_1)(1-t_2) \\
        p-1 & \text{if } 1-t_1 \leq \frac{1}{1+d}(1-s_1)(1-t_2).
    \end{cases}
\end{equation*}
Using this in the calculation of the protocol revenue, yields
\begin{equation*}
    \mathrm{rev}_1 = t_1 (s_1 + (1-s_1-s_2)l_1) = \begin{cases}
        t_1 & \text{if } 1-t_1 \geq \frac{1}{1+d}(1-s_1)(1-t_2) \\
        \frac{s_1(1-t_2)}{(1+d)-(t_2+d)/t_1} & \text{if } 1-t_1 \leq \frac{1}{1+d}(1-s_1)(1-t_2).
    \end{cases}
\end{equation*}

\section{Optimal Trade Calculations}\label{app:optimal_trade}

In the following, we briefly discuss, how trades are routed optimally through multiple AMM pools following considerations from \cite{fritsch21optimalfees}.

The trade function of a constant product market maker with reserves $(A,B)$ and a fee of $f$ is
\begin{equation}\label{eq:trade_function}
    t(x) = B - \frac{AB}{A+(1-f)x}.
\end{equation}
This means that a trader reives $t(x)$ units of the second asset in return for $x$ units of the first asset.

We consider $n$ such pools with reserves $(A_i,B_i)$ and a fee of $f_i$ for $i=1,\ldots,n$, respectively. Each pool has a trade function as in \eqref{eq:trade_function}.
When calulating the optimal trade, we do not consider blockchain transaction fees which can be higher when trading with more pools.
To find the optimal trade, we need to solve

\begin{alignat}{3}
    &\text{maximize} & & t_1(x_1)+\ldots + t_n(x_1) \notag\\
    &\text{subject to} &\quad & x_1+\ldots +x_n = T \tag{OTP}\label{eq:OTP}\\
    & & & x_1, \ldots,x_n \geq 0\notag
\end{alignat}

In our case, it suffices to solve the problem ignoring the constraint $x_1, \ldots,x_n \geq 0$ since it is always optimal to split a trade across all pools if all fees are equal: The assumption that the pools are balanced before the trade implies that the marginal price of an unused pool is always strictly lower than that of a used pool.
Hence, moving an arbitrarily small amount to an unused pool leads to a better trade.

When ignoring the constraint, \eqref{eq:OTP} can be solved using a Lagrange multiplier. This yields the equations $t_i'(x_i)-\lambda = 0$ for $i=1,\ldots,n$ in addition to $x_1+\ldots +x_n = T$.
The former equality implies
\begin{equation*}
    \frac{A_i B_i}{(A_i+(1-f_i)x_i)^2}(1-f_i)-\lambda = 0.
\end{equation*}
Solving this for $x_i$ yields
\begin{equation}\label{eq:x_i_lagrange}
    x_i = \sqrt{\frac{A_i B_i}{1-f_i}}\frac{1}{\sqrt{\lambda}} - \frac{A_i}{1-f_i}.
\end{equation}
Summing this over all $i$ leads to
\begin{equation*}
    T = \left( \sum_{j=1}^n \sqrt{\frac{A_j B_j}{1-f_j}}\right)\frac{1}{\sqrt{\lambda}} - \sum_{j=1}^n \frac{A_j}{1-f_j}.
\end{equation*}
By solving this equation for $1/\sqrt{\lambda}$ and inserting the result back into \eqref{eq:x_i_lagrange}, we conclude
\begin{equation}\label{eq:x_i}
    x_i = \sqrt{\frac{A_i B_i}{1-f_i}}\frac{\left( \sum_{j=1}^n \frac{A_j}{1-f_j}\right)+ T}{\sum_{j=1}^n \sqrt{\frac{A_j B_j}{1-f_j}}} - \frac{A_i}{1-f_i}.
\end{equation}

The assumption of balanced pools implies $A_i/B_i=A_j/B_j$ for all $i,j\in [1,n]$.
That means that we can w.l.o.g. choose the unit of measurement of one of the tokens such that $A_i=B_i$ holds for all pools.
Furthermore, we assumed that trading fees are equal in all pools, i.e.\ $f_i=f_j$ for all $i,j\in [1,n]$.
With these two facts, the term in \eqref{eq:x_i} simplifies to $x_i = (A_i/\sum_{j=1}^n A_j)t$.
Hence, for balanced CFMM pools with identical fees, it is optimal to split the trades proportional to the pool sizes.

\end{document}